\documentclass[twocolumn,aps,prb,10pt,showpacs,raggedbottom,nobalancelastpage,amssymb,superscriptaddress,floatfix]{revtex4-1}
\usepackage{amsmath,amssymb}
\usepackage{txfonts}
\usepackage{hyperref}
\usepackage{braket}
\usepackage{graphicx}
\usepackage{hyperref}

\hypersetup{colorlinks=true, linkcolor=blue, citecolor=blue, urlcolor=blue}

\DeclareMathOperator{\ii}{i\!}
\DeclareMathOperator{\e}{e}
\DeclareMathOperator{\dd}{d\!}
\DeclareMathOperator{\tr}{tr}
\renewcommand{\Re}{\operatorname{Re}}
\renewcommand{\Im}{\operatorname{Im}}

\newcommand\numb{\addtocounter{equation}{1}\tag{\theequation}}
\newcommand{\cH}{\mathcal{H}}

\begin{document}

\title{Lifting the Franck-Condon blockade in driven quantum dots}
\author{Patrick Haughian}
\affiliation{Physics and Materials Science Research Unit, University of Luxembourg, 1511 Luxembourg, Luxembourg}
\author{Stefan Walter}
\affiliation{Institute for Theoretical Physics, University Erlangen-N\"urnberg, Staudtstra\ss e 7, 91058 Erlangen, Germany}
\author{Andreas Nunnenkamp}
\affiliation{Cavendish Laboratory, University of Cambridge, Cambridge, CB3 0HE, United Kingdom}
\author{Thomas L. Schmidt}
\email{thomas.schmidt@uni.lu}
\affiliation{Physics and Materials Science Research Unit, University of Luxembourg, 1511 Luxembourg, Luxembourg}

\date{\today}

\begin{abstract}
Electron-vibron coupling in quantum dots can lead to a strong suppression of the average current in the sequential tunneling regime. This effect is known as Franck-Condon blockade and can be traced back to an overlap integral between vibron states with different electron numbers which becomes exponentially small for large electron-vibron coupling strength. Here, we investigate the effect of a time-dependent drive on this phenomenon, in particular the effect of an oscillatory gate voltage acting on the electronic dot level. We employ two different approaches: perturbation theory based on nonequilibrium Keldysh Green's functions and a master equation in Born-Markov approximation. In both cases, we find that the drive can lift the blockade by exciting vibrons. As a consequence, the relative change in average current grows exponentially with the drive strength.
\end{abstract}

\maketitle

\section{Introduction}
The field of nanoscale electronics has seen rapid advances in recent years: experimental techniques have improved to the point that the range of realizable electronic components now extends down to the single-molecule scale.\cite{rei02, Park2002, Liang2002, Tao2006} Novel fabrication methods afford an increasing amount of precision with regard to the properties of such elements, in particular the conductive behavior. An ultimate goal of this effort is to scale down electronic components such as wires, transistors, and rectifiers to the atomic scale, thus potentially extending the lifetime of Moore's law. Moreover, the physics of nanoscale conductors is not limited to electronic effects. Already at the nanoscale the quantized mechanical degrees of freedom of, e.g., a molecule become important. However, it remains difficult to exploit the mechanical properties of such molecules to control transport through the molecule. Interestingly, the situation is different at the mesoscale, where the interplay between the electronic and mechanical degrees of freedom can be engineered in a fashion that allows for the incorporation of mesoscopic constituents into a wide variety of setups.\cite{wal04,zan12,fre12,xia13}

Suspended carbon nanotubes (CNTs) which are free to vibrate comprise exactly such mesoscopic electromechanical systems.\cite{zan12} CNTs are superb mechanical oscillators due to $(i)$ their high Q-factors and stiffness,\cite{hue09, hue10} $(ii)$ high vibrational frequencies in the $\rm{GHz}$ range,\cite{cha11} and $(iii)$ large electron-phonon coupling.\cite{leturcq2009franck} Besides these mechanical properties, the electronic and transport properties of CNTs can also be tuned depending on the setup. For instance, electronic back gates allow for a controlled shaping of the nanotube's electrostatic potential which can be used to confine single electrons on the CNT, thus creating a quantum dot on the nanotube. Transport through CNT quantum dots has extensively been studied theoretically and experimentally.\cite{Ila2010,Laird2015} Nanoelectromechanics is a growing field of research with various experiments investigating the interplay between the mechanical and the quantized electronic degree of freedom in suspended CNTs.\cite{sap06, ste09, las09, leturcq2009franck}

A general feature of interacting systems composed of electrons and quantized mechanical vibrations (``vibrons'') is the suppression of conductance, in certain parameter regimes, for strong coupling between the two degrees of freedom. This effect, commonly referred to as Franck-Condon blockade,\cite{koc06} results from the atomic constituents of the system accommodating for the presence of a number of electrons by means of displacement, thus forming composite electron-vibron particles termed polarons. Electronic transport through the system requires the electron number to change and hence the polarons to be broken up, which is energetically disfavored if the electron-vibron coupling that holds them together is strong. This effect has been observed in single-molecule junctions\cite{bur14} as well as in CNT systems,\cite{leturcq2009franck} adding to the variety of ways in which material structure can influence conductance. In addition, it has also been shown that the coupling between the mechanical and electronic degrees of freedom can be tailored to some extent.\cite{benyamini2014real}

In this paper we examine the effect of time-dependent driving on the Franck-Condon blockade by periodically modulating the electronic level energy using a time-dependent gate voltage. More specifically, we study a system composed of a suspended CNT on which a quantum dot is defined by means of back gates. This quantum dot is considered weakly coupled to a pair of metallic leads in the regime of sequential tunneling. Moreover, electrons on the dot interact strongly with the vibrational degree of freedom of the CNT. Our goal is to investigate the consequences of periodically modulating the electronic level. Most importantly, we find that driving the system results in a strong increase in the time-averaged current in a way which is reminiscent of a transistor.

\begin{figure}[!t]
\begin{center}
	\includegraphics[width=\columnwidth]{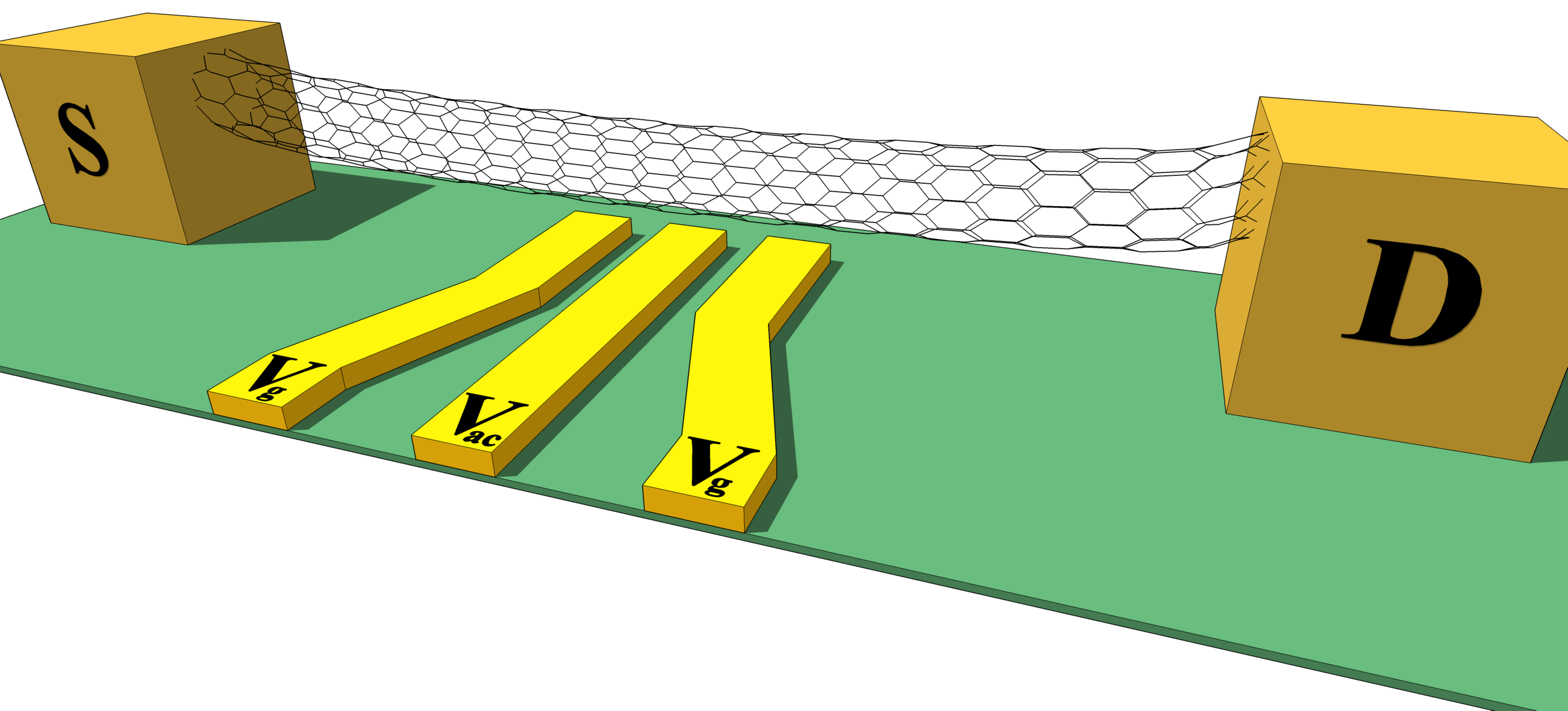}
	\caption{Schematics of the setup: Suspended CNT connected to source (S) and drain (D) electrodes, on top of gate electrodes used to create a quantum dot (\(V_\text{g}\)) and to provide a drive voltage (\(V_\text{ac}\)).}
\label{fig:setup}
\end{center}
\end{figure}

The paper is organized as follows: In Sec.~\ref{sec_model} we present the model used to describe a CNT quantum dot, taking into account a periodic modulation of the electronic level and strong coupling of charge to vibrations. In Sec.~\ref{sec_greens} we derive the steady-state current through the system using the Keldysh nonequilibrium Green's function formalism. Taking an alternative approach, we set up a master equation for the electronic dynamics in Sec.~\ref{sec_master}, leading to a prediction for the current, which we compare to the results presented in Sec.~\ref{sec_greens}. Finally, we summarize in Sec.~\ref{sec_discussion} and conclude by discussing possible applications.

\section{Model}
\label{sec_model}
We consider a quantum dot consisting of a single electronic level weakly tunnel-coupled to a pair of metallic leads. Such a quantum dot may be realized on a suspended CNT using electronic back gates to confine an electron in a specific section of the nanotube.\cite{sap06, ste09, las09, hue09, leturcq2009franck, hue10, cha11, zan12, benyamini2014real} The vibrations of the CNT can be strongly coupled to the charge degree of freedom of the electron and thus have a great influence on its conductive properties.\cite{leturcq2009franck,benyamini2014real,fle06} Additionally, a back gate can be used to apply an ac voltage, thus modulating the dot energy level.\cite{sazonova2004tunable} We show a schematic representation of the setup in
Fig.~\ref{fig:setup}.

This setup can be described by the Anderson-Holstein Hamiltonian, \(\cH=\cH_\text{dot}+\cH_{\text{lead}}+\cH_{\text{tun}}\), where
\begin{align*}
\label{hamiltonian}
\cH_\text{dot}&=\Omega a^{\dagger}a+\epsilon d^{\dagger}d+\lambda(a^{\dagger}+a)d^{\dagger}d+f(t)d^{\dagger}d,\\
\cH_\text{lead}&=\sum_{\alpha=\text{L}, \text{R}} \sum_{k} \omega^{\phantom\dagger}_{k,\alpha}c^{\dagger}_{k,\alpha}c^{\phantom\dagger}_{k,\alpha},\\
\cH_\text{tun}&=g\sum_{\alpha=\text{L}, \text{R}}\left[d\psi^\dagger_{\alpha}(x=0)+\text{h.c.}\right],\numb
\end{align*}
denote the dot, lead, and tunneling Hamiltonians, respectively.
\begin{figure}[!ht]
\begin{center}
	\includegraphics[width=\columnwidth]{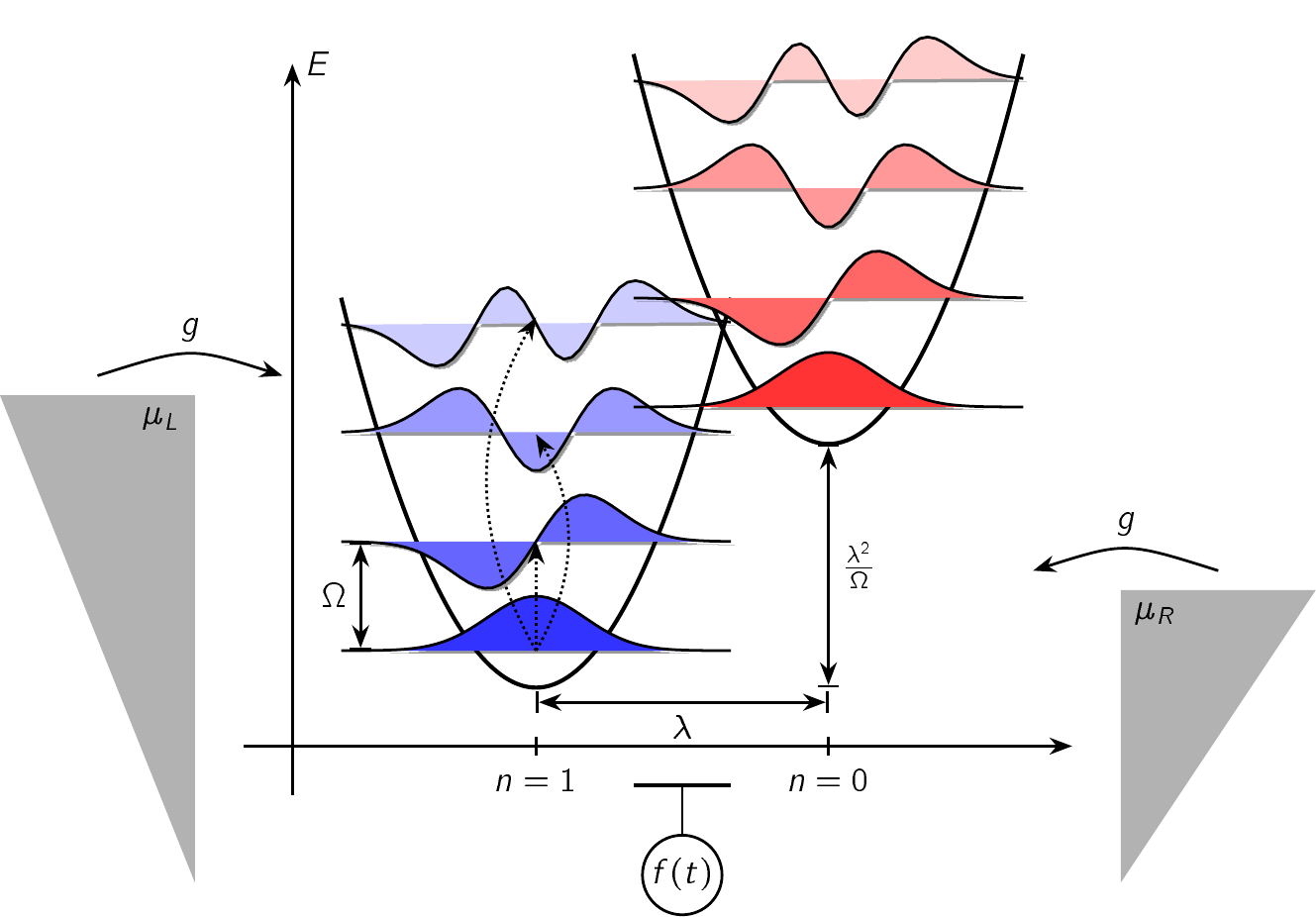}
	\caption{Illustration of the Hamiltonian from Eq.~\eqref{hamiltonian}. Electron-vibron coupling corresponds to a shift of the vibron rest position depending on the electron number $n = d^\dag d$. This shift leads to an exponentially small overlap between the oscillator's ground state wave functions for $n=0$ and $n=1$. However, driving [\(f(t)\)] causes transitions to excited states (dashed arrows) for which the overlap is significantly larger. Changes in electron number occur by tunneling into either of the leads at chemical potentials \(\mu_\text{L}\) and \(\mu_\text{R}\), respectively, with tunneling amplitude \(g\).}
\label{fig:setupsketch}
\end{center}
\end{figure}

While in principle a CNT admits several different types of vibron modes, the coupling to the charge sector is strongest for breathing and longitudinal stretching modes.\cite{vop09} In our analysis, we restrict the model for the quantum dot to a single electronic level at energy \(\epsilon\) and a single vibron mode of frequency \(\Omega\), setting \(\hbar=1\) throughout. In addition, we include the gate-induced drive as a time-dependent contribution \(f(t)\) to the electronic level energy. The mechanical vibration of the nanotube modulates the dot level energy, which is quantified by the coupling strength \(\lambda\). The leads $\alpha \in \{\text{L}, \text{R}\}$ are modeled as fermionic reservoirs with single-particle energies \(\omega_{k,\alpha}\), described by fermionic operators \(c_{k, \alpha}\) obeying the canonical anticommutation relations $\{ c^{\phantom\dagger}_{k, \alpha}, c^{\dag}_{k', \beta} \} = \delta_{k,k'}\delta_{\alpha,\beta}$. Finally, \(\cH_\text{tun}\) describes local electron tunneling into and out of the leads using the Fourier transform $\psi_\alpha(x) = L^{-1/2}\sum_{k} e^{\mathrm{i} k x} c_{k,\alpha}$, where $L$ is the length of the lead. An illustration of the different terms of \(\mathcal{H}\) is shown in Fig.~\ref{fig:setupsketch}.

In the absence of electron-vibron coupling, the dot features a single resonance at energy $\epsilon$. Electron-vibron coupling leads to the emergence of side-peaks at energies $\epsilon + n \Omega$ with $n \in \mathbb{Z}$. In the following, we focus on the limit of sequential tunneling. This is the dominant transport process for small $g$ and potential differences \(V = \mu_{\text{L}} - \mu_{\text{R}}\), where $\mu_{\alpha}$ is the chemical potential of lead $\alpha$, such that only a single resonance lies within the bias window.

The electron-vibron coupling term can be removed by applying the polaron transformation\cite{lf63}  given by \(\mathcal{U}=\exp[\lambda(a^{\dagger}-a)d^{\dagger}d/{\Omega}]\), leading to
\begin{align*}
&\mathcal{U}\cH\mathcal{U}^{-1} = \Omega a^{\dagger}a+\tilde{\epsilon} d^{\dagger}d+f(t)d^{\dagger}d \\
	&+\sum_{\alpha=\text{L}, \text{R}}\sum_{k}\omega^{\phantom\dagger}_{k,\alpha}c^{\dagger}_{k,\alpha}c^{\phantom\dagger}_{k,\alpha}
+g\sum_{\alpha=\text{L}, \text{R}}\left[X^\dagger d\psi_\alpha^\dagger(x=0)+\text{h.c.}\right] \numb,
\end{align*}
where the electron level energy \(\epsilon\) is renormalized to \(\tilde{\epsilon}=\epsilon-\lambda^2/\Omega\) and the electron-vibron coupling is moved to an exponential factor multiplying the tunneling term, \(X=\mathrm{e}^{-\lambda(a^{\dagger}-a)/{\Omega}}\). This leaves us with an expression consisting of a quadratic Hamiltonian $\cH_0$ and a weak perturbation $\cH_1$,
\begin{align*}
\cH_0 &= \Omega a^{\dagger}a+\tilde{\epsilon} d^{\dagger}d+f(t)d^{\dagger}d+\sum_{\alpha}\sum_{k}\omega^{\phantom\dagger}_{k,\alpha}c^{\dagger}_{k,\alpha}c^{\phantom\dagger}_{k,\alpha} \notag \\
\cH_\text{I} &= g\sum_{\alpha}\left[X^\dagger d\psi_\alpha^\dagger(x=0)+\text{h.c.}\right],\numb
\end{align*}
which lends itself to a variety of approaches that are perturbative in the tunneling amplitude \(g\), but still permit potentially large values of the electron-vibron coupling \(\lambda\gg\Omega\).

A great deal of insight into similar models has already been obtained. Specifically, the undriven \([f(t)=0]\) variant of the system has been examined with regard to its transport properties.\cite{koc04,koc05,koc06,sch09,riw09} The most striking finding in this context is that of Franck-Condon blockade: strong electron-vibron interaction leads to formation of a composite state, called a polaron, which can be thought of as being made up of an electron and a ``cloud'' of vibrational excitations surrounding it. If the electron is to tunnel out of the quantum dot, this state has to be broken up, at an energy cost which strongly increases with the coupling, leading to an exponential suppression of tunneling.

In the following, we will investigate in detail the novel effects that arise by periodically modulating the electronic level energy. We will first show that due to the strong electron-vibron coupling, this type of drive can be mapped to a drive of the vibron. Moreover, we will demonstrate that it has a strong influence on electronic transport and that, in particular, the Franck-Condon blockade can be lifted. It is worth pointing out that a small ac drive voltage applied to the gate can lead to an exponentially strong change of the average current.

The physical interpretation of this process is as follows. Due to the electron-vibron coupling, a time-dependent gate voltage has the same effect as driving the vibron, so the vibron will populate an excited state. Importantly, the overlap integral between excited vibron states for different fermion numbers contains Franck-Condon factors which are exponentially larger than those of the vibron ground states. This makes it possible to lift the Franck-Condon blockade, and hence increases the average current.

Hence, the relative change in average current $I(A)/I(0)$, where $A$ is the amplitude of the ac gate voltage, is close to an exponential function of $A$. This makes the $I-V$ characteristic reminiscent of that of a transistor, albeit now for ac gate voltages: in our system, a dc bias current can be controlled by a small ac gate voltage. Note that the frequency of the ac gate voltage needs to be close to the vibron frequency, since the effect rests on the excitation of vibron.

\section{Polaron tunneling approximation}
\label{sec_greens}
In the following, we use a Keldysh Green's function approach to analyze the influence of the gate voltage on the conductance of the system. The most important measurable quantity in this context is the steady-state current \(\langle I \rangle \). In the following, we work with the current through the right lead, noting that it is identical to the one through the left lead up to a displacement current introduced by the time dependence of the drive. However, since this current oscillates on a timescale given by the inverse drive frequency, much faster than the tunneling rates to and from the dot, we can safely neglect it on average. Thus \(\langle I\rangle\) is given by the expectation value of the operator
\begin{align*}
\label{currentop}
I&=-e\frac{d}{dt} \sum_{k}c^{\dagger}_{k,\text{R}}c^{\phantom\dagger}_{k,\text{R}} = \mathrm{i} e \left[\sum_{k}c^{\dagger}_{k, \text{R}}c^{\phantom\dagger}_{k,\text{R}}, \cH \right] \\
&= -\mathrm{i}eg\left[d^{\dagger}(t)X(t)\psi^{\phantom\dag}_\text{R}(t)-\psi^\dagger_\text{R}(t)X^{\dagger}(t)d(t)\right],\numb
\end{align*}
where $e$ is the negative electron charge. It will prove expedient to move the time dependence into the perturbative part of the Hamiltonian, which can be accomplished by applying the unitary transformation given by \(\mathcal{V}(t)=\exp\big[\mathrm{i} d^{\dagger}d\int_{t_0}^{t}\mathrm{d} sf(s)\big]\). This reduces the quadratic Hamiltonian to \(\bar{\mathcal{H}}_0=\tilde{\epsilon}d^\dagger d+\Omega a^\dagger a\), and changes the polaron operator to
\begin{align}
\bar{X}(t)\equiv\mathcal{V}(t)X(t)\mathcal{V}^{-1}(t)=\mathrm{e}^{-\frac{\lambda}{\Omega}\left[a^{\dagger}\mathrm{e}^{\mathrm{i}\Omega t}-a\mathrm{e}^{-\mathrm{i}\Omega t}+\mathrm{i} F(t)\right]},
\end{align}
with \(F(t)=\frac{\Omega}{\lambda}\int_{t_0}^{t}\mathrm{d} sf(t)\) denoting the integrated drive, where the initial time \(t_0\) always cancels in the following and can hence be chosen arbitrarily. The transformed coupling Hamiltonian then reads 
\begin{align}
\bar{\mathcal{H}}_\text{I}=g\sum_{\alpha}\left[\bar{X}^\dagger d\psi_\alpha^\dagger(x=0)+\text{h.c.}\right].
\end{align}
It bears pointing out that as a result of the transformation \(\mathcal{V}\) the drive has thus been moved onto the vibrational part of the Hamiltonian, lending substance to the intuition that the coupling between electrons and vibrons leads to the possibility of driving the vibrons by driving the electrons.

In the following, we focus on a resonant harmonic drive, \({f(t)=A\cos(\Omega t)}\). The Keldysh Green's function formalism together with perturbation theory in the lead coupling \(g\) can be employed to calculate the mean steady-state current\cite{mw92} from the retarded Green's function \(D^\text{R}\),
\begin{align}
\label{currentexp}
\langle I \rangle =&-\mathrm{i}\frac{e}{\pi}\frac{\Gamma}{4}\int\dd\omega\left[f_{\text{L}}(\omega)-f_{\text{R}}(\omega)\right]\left[D^{+-}(\omega)-D^{-+}(\omega)\right],
\end{align}
where the coupling $g$ and the lead density of states \(\nu\) are absorbed into the tunneling rate \(\Gamma=2\pi\nu g^2\). We used the wide-band approximation, so the density of states is constant, $\nu = 1/(2 \pi v_\text{F})$ with Fermi velocity $v_\text{F}$. Hence, to obtain the average current we need to calculate the Keldysh dot Green's function 
\begin{align}
\label{dotgf}
D(t,t^\prime)=-\mathrm{i}\left\langle\mathcal{T}_{C}d^{\dagger}(t)\bar{X}(t)\bar{X}^{\dagger}(t^\prime)d(t^\prime)\mathrm{e}^{-\mathrm{i}\int_{t^{\prime}}^{t}\mathrm{d} s\bar{\cH}_{\text{I}}(s)}\right\rangle_0,
\end{align}
with components \(D^{ij}(t,t^\prime)\), where the indices \(i,j=+,-\) denote the forward and backward branches of the Keldysh contour running from \(-\infty\) to \(\infty\) and back again, respectively. The symbol \(\mathcal{T}_C\) indicates path-ordering along the Keldysh contour. The expectation value in Eq.~(\ref{dotgf}) is taken with respect to the Hamiltonian \(\bar{\cH}_0\) of the non-interacting system. We note that the presence of the vibrational operators \(\bar{X}\) in this expression is a consequence of the polaron transform dressing the electron. As the non-interacting Hamiltonian is not quadratic in \(\bar{X}\), \(\bar{X}^\dagger\), Wick's theorem does not hold for these operators, so the calculation of the terms making up the Green's function in Eq.~\eqref{dotgf} seems rather daunting.

\begin{figure}[t]
     \includegraphics[width=0.99\columnwidth]{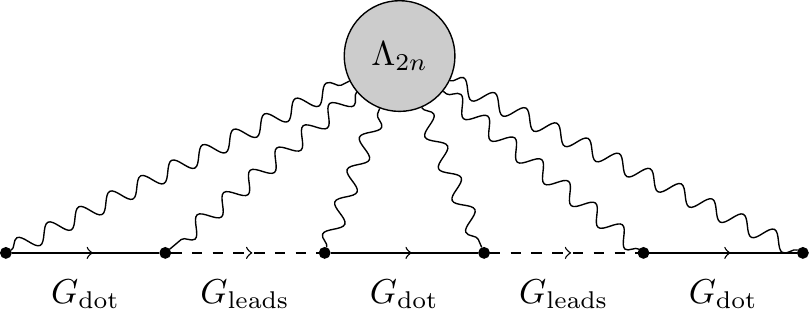}
     \includegraphics[width=0.99\columnwidth]{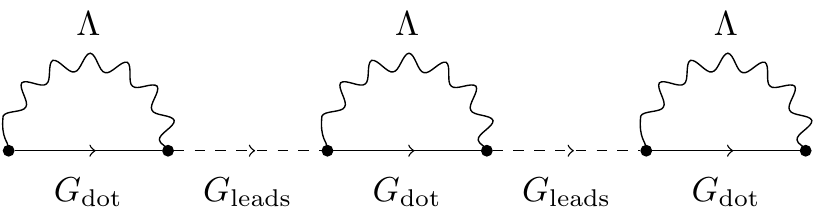}
 \caption{\emph{Top}: Generic diagram in the perturbation expansion of the full dot propagator, with \(\Lambda_{2n}\) denoting the \(2n\)-correlator of \(\bar{X}\) operators, bare dot propagator \(G_{\text{dot}}\), and lead propagator \(G_{\text{leads}}\). \emph{Bottom}: Polaron tunneling approximation of the same diagram. The vibron cloud is assumed to de-excite between electron tunneling processes, leaving only correlators of second order in the vibron sector, i.e.~the top diagram reduces to the bottom one.}
\label{PTAdiagrams}
\end{figure}

Drawing upon Ref.~[\onlinecite{Mai11}], we hence employ the following approximation: the dwell time of the electron on the quantum dot (which can be estimated as the inverse of the bare tunneling rate \(\Gamma\)) is large compared to the timescale of the polaron which is associated with the inverse of the energy shift \(\epsilon-\tilde{\epsilon}=\lambda^2/\Omega\). In this limit, the polaron will relax in the time between two tunneling processes. The diagrammatic form of this polaron-tunneling approximation (PTA) is given in Fig.~\ref{PTAdiagrams}, and it has been used before to calculate transport properties of strongly coupled electron-vibron systems.\cite{Mai11,sou14,alb13} The purpose of this approximation is to avoid having to explicitly expand the exponential from Eq.~\eqref{dotgf} in powers of the tunneling amplitude \(g\). Instead, the \(2n\)-th order vibrational correction to the bare dot Green's function \(G_\text{dot}(t-t^\prime) = -i \langle\mathcal{T}_{C} d(t) d^\dagger(t^\prime) \rangle \) is seen to be caused by a series of tunneling processes to and from the leads, where the lead Green's function is given by
\begin{align}
G_{\text{leads}}(\omega)=\mathrm{i}2\pi\nu\begin{pmatrix}n_\text{L}+n_\text{R}-\frac{1}{2}&n_\text{L}+n_\text{R}\\n_\text{L}+n_\text{R}-1&n_\text{L}+n_\text{R}-\frac{1}{2}\end{pmatrix},
\end{align}
where $n_{\text{L}} = n_\text{F}(\omega - eV/2)$ and $n_{\text{R}} = n_\text{F}(\omega + eV/2)$ are the Fermi functions of the left and right lead, respectively. Here, we also made use of the wide band-limit of the lead distributions, as detailed in App.~\ref{sec_app}. Each tunneling process involves the mechanical degree of freedom, giving rise to vibron excitations described by the set of correlators 
\begin{align}
    \Lambda_{2n}(t^{\phantom\prime}_1,t^\prime_1 \ldots,  t^{\phantom\prime}_{n},t^\prime_n)=\left\langle\mathcal{T}_{C}\prod_{1 \leq j\leq n}\bar{X}(t^{\phantom\dag}_j)\bar{X}^\dagger(t_j^\prime)\right\rangle.
\end{align}
The PTA replaces these by products of the quadratic correlators \(\Lambda(t,t^\prime)=\langle \mathcal{T}_{C}\bar{X}(t)\bar{X}^\dagger(t^\prime)\rangle\) connecting only two consecutive tunneling events into and out of the quantum dot. These correlators consist of a drive-independent factor and one which explicitly incorporates the drive, i.e., \(\Lambda(t,t^\prime)=\Lambda_0(t-t^\prime)\Lambda_\text{dr}(t,t^\prime)\), with
\begin{align}
\label{lambda2}
\Lambda_0(t-t^\prime)=-\mathrm{i}\mathrm{e}^{-\frac{\lambda^2}{\Omega^2}}\begin{pmatrix}\mathrm{e}^{\frac{\lambda^2}{\Omega^2}\mathrm{e}^{-\mathrm{i}\Omega\left\vert t-t^\prime\right\vert}}&\mathrm{e}^{\frac{\lambda^2}{\Omega^2}\mathrm{e}^{\mathrm{i}\Omega (t-t^\prime)}}\\\mathrm{e}^{\frac{\lambda^2}{\Omega^2}\mathrm{e}^{-\mathrm{i}\Omega (t-t^\prime)}}&\mathrm{e}^{\frac{\lambda^2}{\Omega^2}\mathrm{e}^{\mathrm{i}\Omega\left\vert t-t^\prime\right\vert}}
\end{pmatrix}
\end{align}
in Keldysh space, and the driven component
\begin{align}
\Lambda_\text{dr}(t,t^\prime)=\mathrm{e}^{-\mathrm{i}\int_{t^{\prime}}^{t}\mathrm{d} sf(s)}.
\end{align}
Since the timescale of the vibrons is fast compared to the tunneling, we average the driven part over one drive period, leading to (see App.~\ref{sec_app})
\begin{align}
\Lambda_\text{dr}(t-t^\prime)=J_0\left(\frac{2A\sin{\Omega(t-t^\prime)/2}}{\Omega}\right),
\end{align}
where \(J_0\) denotes the Bessel function of the first kind. This expression only depends on the relative time coordinate \(t-t^\prime\), meaning that the problem becomes readily amenable to Fourier transformation. This leaves us with the leading-order correction in frequency space to the dot propagator,
\begin{align}
D_0^{ij}(\omega)=\int_{-\infty}^{\infty}\mathrm{d} t \mathrm{e}^{\mathrm{i}\omega t}G_{\text{dot}}^{ij}(t)\Lambda^{ij}(t).
\end{align}

\begin{figure}[t]
\centering
\includegraphics[width=0.99\columnwidth]{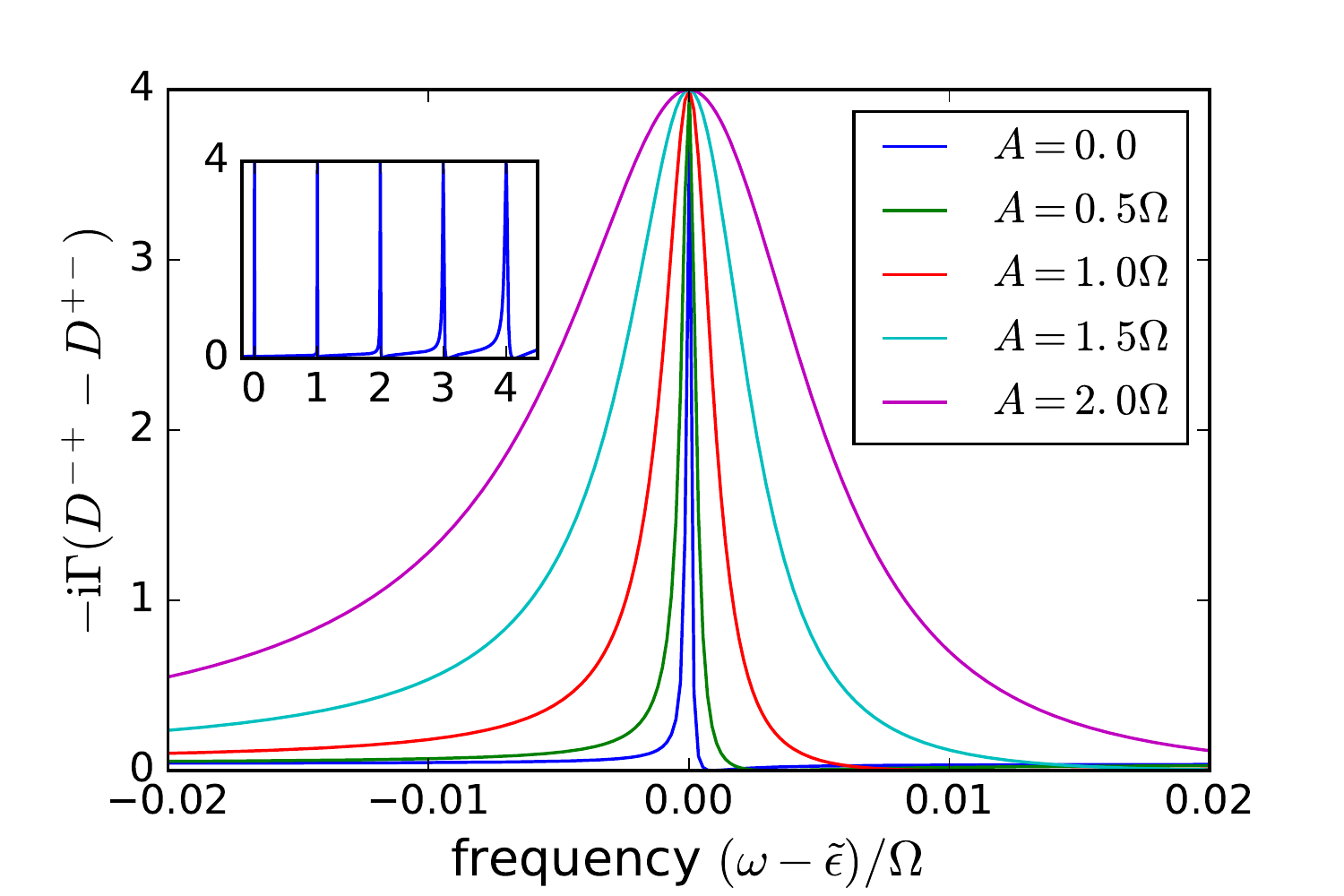}
\caption{Zeroth-order peak (centered around \(\omega=\tilde{\epsilon}\)) of the PTA spectral function \(-\mathrm{i}\left[D^{+-}(\omega)-D^{-+}(\omega)\right]\) for different values of the drive amplitude \(A\). Stronger drive leads to a significant broadening of the peaks. \emph{Inset}: PTA Spectral function, with peaks at every integer multiple of the vibron frequency. Peak width is suppressed by the Franck-Condon factor \(\e^{-\lambda^2/\Omega^2}\), but increases proportionally to \(\Gamma\) and as a function of frequency. The latter is a result of higher favorability of transitions involving large excitations of the vibron mode.}
\label{drivezeropeak}
\end{figure}

At this point, the simplified structure of the diagrams in Fig.~\ref{PTAdiagrams} allows us perform a partial resummation and thus retain all orders in the coupling strength \(g\) without requiring access to higher-order vibrational correlators. Indeed, incrementing the order in the tunneling is equivalent to appending a single copy each of the polaron-dot and lead propagators to the diagram. In terms of Green's functions, this is equivalent to the frequency-space Dyson equation for the vibron-dressed dot electron propagator,
\begin{align}
\label{dyson}
D^{-1}(\omega)=D_0^{-1}(\omega)-\Sigma_{\text{leads}}(\omega).
\end{align}
with the self-energy \(\Sigma_\text{leads}=g^2G_\text{leads}\). This relation gives rise to the full PTA dot Green's function \(D(\omega)\), as documented in App.~\ref{sec_app}. The steady-state current from Eq.~(\ref{currentexp}) can now be obtained by integrating over the spectral function \(-\mathrm{i}\left[D^{-+}(\omega)-D^{+-}(\omega)\right]\), which is shown in Fig.~\ref{drivezeropeak}. The plot features tunnel-broadened resonances at all integer multiples of the vibron frequency. The appearance of peak heights being independent of peak order is an artifact caused by the PTA;\cite{sou14} only the zeroth-order peak is reliable in that regard. Furthermore, the electron-vibron coupling causes additional broadening of the peaks, which grows larger as a function of peak order. The ac drive causes the peak widths to increase strongly as a function of the drive amplitude, as illustrated in Fig.~\ref{drivezeropeak}.

With the Green's function in Eq.~\eqref{dyson} we can calculate the steady-state current through the quantum dot
\begin{align}
\label{current}
\langle I \rangle = -\frac{e}{\pi}\frac{\Gamma}{4}\int_{-eV/2}^{eV/2}\dd\omega\frac{2\Gamma}{\det{D^{-1}(\omega)}},
\end{align}
with \(V\) denoting the finite bias voltage between left and right lead, where we take the limit of zero temperature in the leads, $T_{\text{el}}=0$. The determinant \(\det{D^{-1}(\omega)}\) entails taking the sum over all vibron resonances, modified by the drive. The sequential-tunneling regime is characterized by the fact that only a single such resonance lies within the bias window, i.e., the vibron frequency is much larger than the bias \(eV\). Moreover, the width of the resonances is proportional to the square of the tunneling amplitude and suppressed by the Franck-Condon factor, i.e., the resonance width is small compared to the bias window, see Fig.~\ref{drivezeropeak}. Taken together, this allows us to consider only a single resonance as integrand, and to expand the limits of integration in Eq.~\eqref{current} to infinity, resulting in the analytic expression for the current
\begin{align}
\langle I \rangle = 2e\Gamma\mathrm{e}^{-\frac{\lambda^2}{\Omega^2}}\sum_{p\in\mathbb{N}}\Lambda_{\text{dr}}^{(-p)}\frac{1}{p!}\left(\frac{\lambda^2}{\Omega^2}\right)^p,
\end{align}
where \(\Lambda_{\text{dr}}^{(-p)}\) denotes the \((-p)\)th Fourier coefficient of the drive component in \(\Lambda\),
\begin{align}
\Lambda_{\text{dr}}^{(n)}=\sum_{m\geq |n|}\frac{(-1)^{m-n}}{m!m!}\left(\frac{A}{2\Omega}\right)^{2m}\begin{pmatrix}2m\\m+n\end{pmatrix} .
\end{align}
For details we refer to App.~\ref{sec_app}.

\begin{figure}[t]
	\includegraphics[width=0.99\columnwidth]{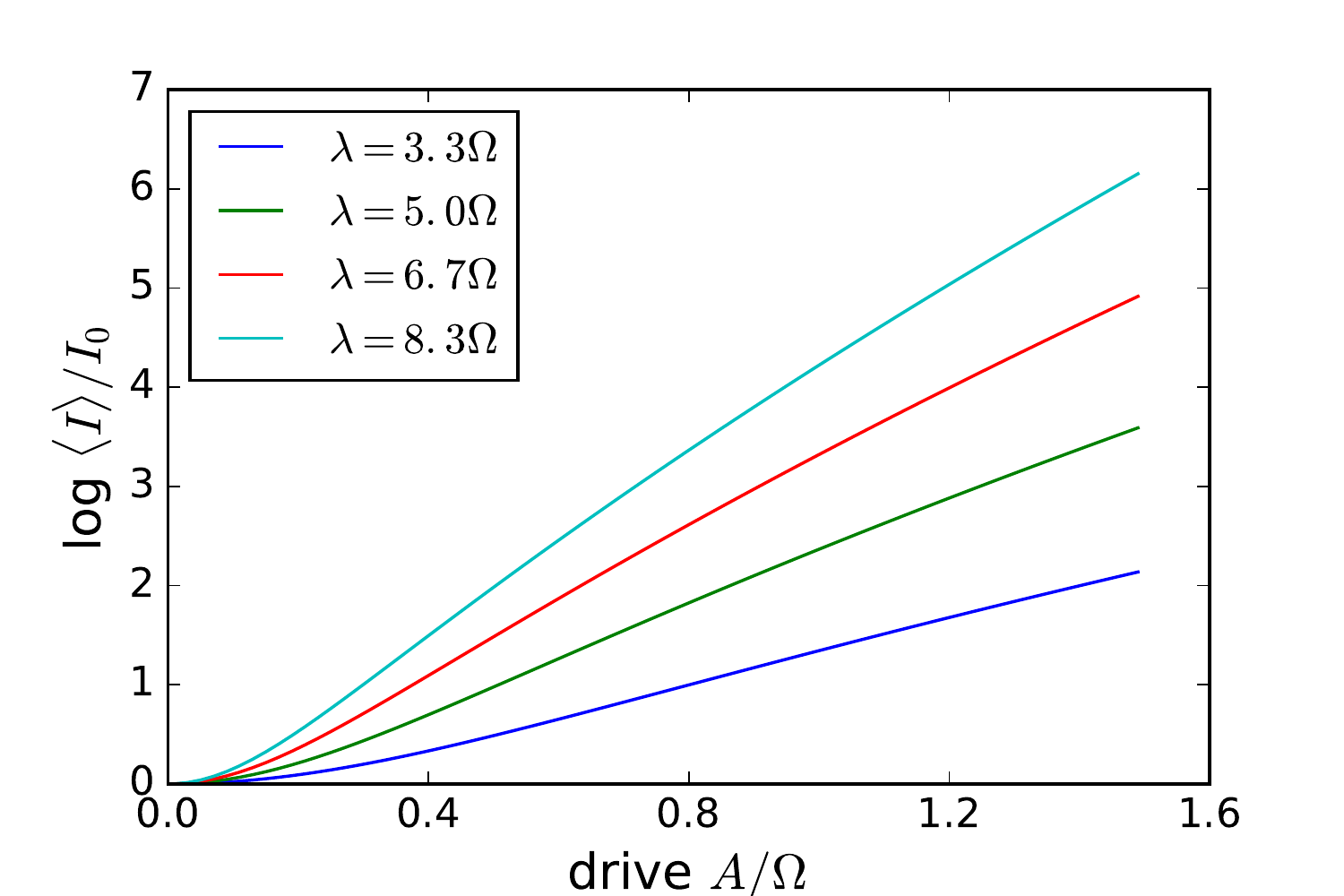}
\caption{Current (logarithmic scale) through the resonantly driven dot, scaled by \(I_0=\e^{-\lambda^2/\Omega^2}\Gamma\), as a function of driving amplitude \(A\) for different choices of strong electron-vibron coupling \(\lambda\). Drive lifts the blockade in an almost exponential fashion, with stronger response for larger coupling.}
\label{drivecurrent}
\end{figure}

As shown in Fig.~\ref{drivecurrent}, the current increases almost exponentially 
as a function of the drive amplitude \(A\), indicating that the Franck-Condon blockade can be lifted through a periodic modulation of the gate voltage. It is to be noted that the deviation from an exponential behavior grows larger as the drive becomes stronger, which might point towards a progressive failure of the PTA. Nonetheless, the above results provide substantial evidence that a CNT quantum dot in the sequential-tunneling regime with strong electron-vibron coupling can be turned into a tunable conductor by means of a time-dependent gate voltage.

\section{Born-Markov analysis}
\label{sec_master}
In this section we make use of an alternative approach to the setup under consideration, thus supplementing the result of the Green's function calculation. In particular, we aim to obtain an estimate of the validity the PTA, whose compatibility with a time-dependent driving of the electronic level energy is still untested. For the purposes of the following discussion, it suffices to take into account a single lead, and instead of a steady-state current, study the tunneling rates into this lead, which in the limit of sequential tunneling is proportional to the conductance.\cite{koc06} The comparability of the latter two quantities is a feature of the sequential-tunneling regime, where the current is made up of consecutive, non-overlapping single-electron tunneling transitions.

The system being weakly coupled to a metallic lead with intractably many degrees of freedom allows us to perform a partial trace over the lead degrees of freedom of the model. This well-established approach treats the quantum dot as an open quantum system and the lead as an electron reservoir.\cite{tim08} The weak coupling Hamiltonian \(\cH_{\text{I}}\) then takes the role of a perturbative system-bath interaction. To second order in this interaction, the evolution of the system's density matrix \(\rho_\text{S}\)(t) is governed by the quantum master equation
\begin{align}
\label{redf}
\dot{\rho}_{\text{S}}(t)=&-\int_{0}^{\infty}\mathrm{d} s\tr_{\text{B}}[\cH_\text{I}(t),[\cH_\text{I}(t-s),\rho_{\text{S}}(t)\otimes\rho_{\text{B}}]],
\end{align}
where the time-dependence of the Hamiltonian is understood with respect to the quadratic Hamiltonian \(\Omega a^{\dagger}a+[\tilde{\epsilon}+f(t)]d^{\dagger}d\), and \(\rho_\text{B}\) denotes the bath density matrix. In order to obtain the above expression, we have performed a Born-Markov approximation. Specifically, the total density matrix is presumed to factor into a system and a bath component, and the bath degrees of freedom are taken to relax much faster than those of the system. At this point, an additional simplification presents itself as a consequence of the configuration of our quantum dot: in the Coulomb blockade and sequential tunneling regimes not only the lead excitations, but also the vibrations have fast dynamics compared to the electron on the quantum dot. This means that after the polaron transform the non-perturbative part of the Hamiltonian no longer couples the charge and vibrational degrees of freedom. Hence we can extend the bath to also include the vibron, leaving only the electron in the system.\cite{kra15,sch13} This leads to the bath density matrix composed of the vibrons and the lead \(\rho_\text{B}=\rho_\text{vib}\otimes\rho_\text{lead}\). Note that the separation of time scales between electron and vibron relaxation is similar to the reasoning used to motivate the PTA in the previous section. Nevertheless, the two approximations correspond to different partial resummations of a perturbation series and should not be expected to coincide quantitatively.

The Born-Markov approximation often gives rise to a master equation of Lindblad form, meaning that a set of coefficients \(h_\mu\) and system operators \(C_\mu\) can be found such that
\begin{align}
\label{lindbladops}
\dot{\rho}_{\text{S}}=\sum_{\mu}h_{\mu} \left(C_{\mu}\rho_{\text{S}}C^\dagger_{\mu}-\frac{1}{2} \left\{ \rho_{\text{S}}, C^\dagger_{\mu}C_{\mu} \right\} \right),
\end{align}
which is the most general master equation ensuring the positivity of the density matrix and preserving its trace. As in our case the vibron degree of freedom is traced over, the Lindblad operators \(C_\mu\) will be simple functions of the electronic creation and annihilation operators.

However, the presence of the driving term \(f(t)d^{\dagger}d\) casts doubts upon the validity of the approximations leading to this form: in particular, driving the system with a frequency comparable to \(\Omega\), as is realistic in the CNT setup, is in conflict with the assumption of separation of time scales, now that the bath includes the vibron.
Therefore, we propose the following alternative path to incorporate driving: in a fashion reminiscent of the transformation \(\mathcal{V}\) from the previous section, the driving term can be moved onto the vibron sector, where it then can be taken into account by modifying the vibron density matrix in the master equation \eqref{redf}. Assuming equilibrium Fermi distributions \(n_\text{F}(\omega)\) in the lead, the master equation then reads
\begin{align*}
\label{rhoeq2}
&\dot{\rho}_{\text{S}}(t)=-g^{2}\int_{0}^{\infty}\mathrm{d} s\int\dd\omega \\
&\left[\left\langle\tilde{X}(t)\tilde{X}^{\dagger}(t-s)\right\rangle n_\text{F}(\omega)dd^{\dagger}\mathrm{e}^{\mathrm{i}\left(-\tilde{\epsilon}+\omega\right)s}\rho_{\text{S}}(t)\right.\\
&+\left\langle\tilde{X}^{\dagger}(t)\tilde{X}(t-s)\right\rangle\left(1-n_\text{F}(\omega)\right)d^{\dagger}d\mathrm{e}^{-\mathrm{i}\left(-\tilde{\epsilon}+\omega\right)s}\rho_{\text{S}}(t)\\
&-\left\langle\tilde{X}^{\dagger}(t-s)\tilde{X}(t)\right\rangle d\rho_{\text{S}}(t)d^{\dagger} \left(1-n_\text{F}\left(\omega\right)\right)\mathrm{e}^{\mathrm{i}\left(-\tilde{\epsilon}+\omega\right)s}\\
&-\left.\left\langle\tilde{X}(t-s)\tilde{X}^{\dagger}(t)\right\rangle d^{\dagger}\rho_{\text{S}}(t)dn_\text{F}(\omega)\mathrm{e}^{-\mathrm{i}\left(-\tilde{\epsilon}+\omega\right)s}\right]+\text{h.c.}\numb
\end{align*}
where the expectation values are taken with respect to the vibron density matrix \(\rho_\text{vib}\), which we use below to take into account the drive. Moreover, \(\tilde{X}(t)=\exp\big[-\frac{\lambda}{\Omega}\left(a^{\dagger}\mathrm{e}^{\mathrm{i}\Omega t}-a\mathrm{e}^{-\mathrm{i}\Omega t}\right)\big]\) denotes the exponential vibron operator. We again assume the lead to be at zero temperature, $T_{\text{el}}=0$. In the following, we explore different choices for \(\rho_\text{vib}\), and study the resulting electron tunneling rates to which they lead.

\subsection{Vibron in the ground state}
\label{sec_master1}
\begin{figure}[t]
     \includegraphics[width=0.99\columnwidth]{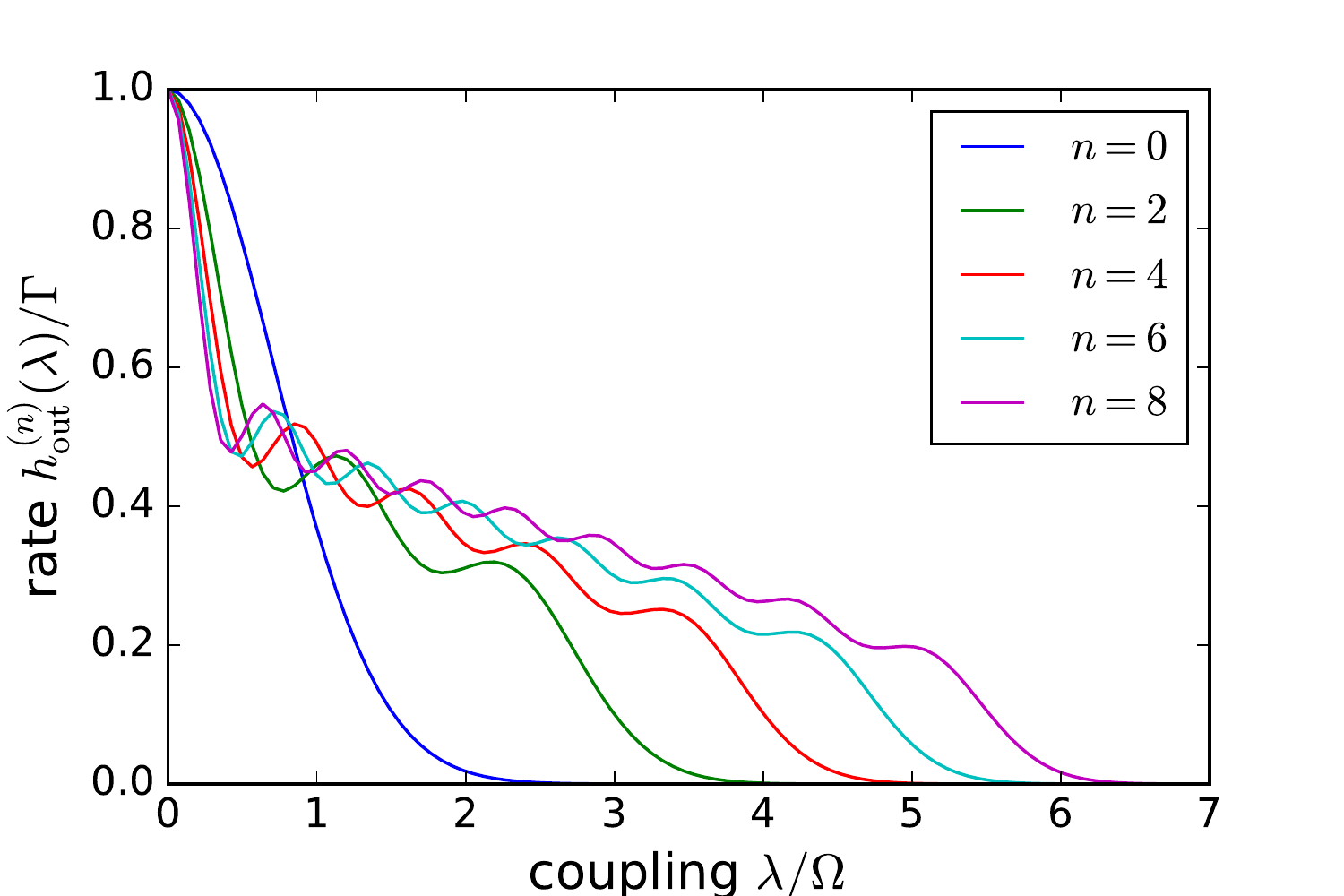}
 \caption{Dot electron tunneling rate into the lead, as a function of electron-vibron coupling \(\lambda\), represented by the Lindblad coefficient \(h^{(n)}_\text{out}\) for different vibron Fock states \(n\), including the ground state \(n=0\) [Eq.~\eqref{lowtemp}]. In the latter case, the rate decreases exponentially for higher electron-vibron coupling, exemplifying Franck-Condon blockade. Higher vibron numbers lead to an intermediate regime of less pronounced decrease. In addition, \(n\) manifests itself as the number of local maxima superimposed over the decaying curve.}
\label{lindbladrates}
\end{figure}

To begin with, we assume an unoccupied vibron state and thus \(\rho_\text{vib}^{(0)}=\ket{0}\bra{0}\), corresponding to the absence of driving and zero temperature. In this case, the traces in Eq.~(\ref{rhoeq2}) are readily performed. Comparing to the master equation in Lindblad form Eq.~(\ref{lindbladops}) leads to the Lindblad operators \(C^{(0)}_\text{in}=c^\dagger\) and \(C^{(0)}_\text{out}=c\) with coefficients
\begin{align*}
\label{lowtemp}
h^{(0)}_\text{in}=& \frac{\Gamma}{2} \mathrm{e}^{-\frac{\lambda^2}{\Omega^2}}\sum_{n\geq0}\frac{\left(\lambda^2/\Omega^2\right)^{n}}{n!}n_\text{F}(\tilde{\epsilon}+n\Omega),\\
h^{(0)}_\text{out}=&\frac{\Gamma}{2} \e^{-\frac{\lambda^2}{\Omega^2}}\sum_{n\geq0}\frac{\left(\lambda^2/\Omega^2\right)^{n}}{n!}\left(1-n_\text{F}(\tilde{\epsilon}-n\Omega)\right).\numb
\end{align*}
Here \(J_n\) denotes the \(n\)th Bessel function of the first kind, and \(\Gamma\) is defined as before. Due to the specific form of the Lindblad operators, the coefficients \(h^{(0)}_\text{in}\) and \(h^{(0)}_\text{out}\) are proportional to the electron tunneling rates. Specifically, the respective appearances of \(n_{F}\) and \(1-n_{F}\) indicate that \(h^{(0)}_\text{out}\) governs the case of tunneling into the lead, whereas \(h^{(0)}_\text{in}\) describes tunneling in the other direction. The resulting dependence of the decay rates on the coupling strength is shown in Fig.~\ref{lindbladrates}, illustrating the Franck-Condon blockade in the strong-coupling regime. More quantitatively, we note that the factor \(\mathrm{e}^{-\lambda^2/\Omega^2}\) matches the Franck-Condon blockade strength obtained from a classical rate equation.\cite{koc06}

\subsection{Vibron in a Fock state}
\label{sec_master2}
\begin{figure}[t]
     \includegraphics[width=0.99\columnwidth]{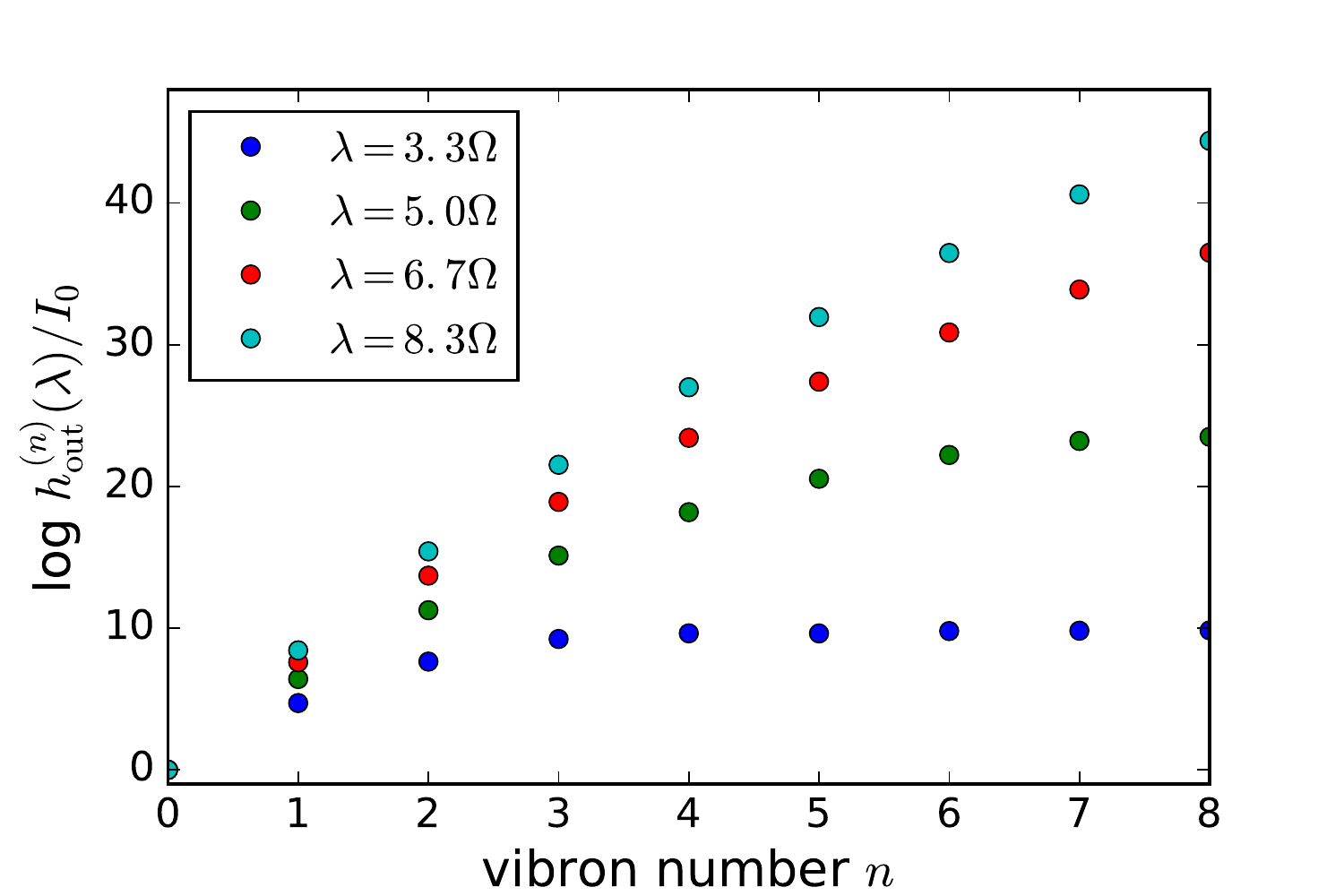}
 \caption{Tunneling rate (logarithmic scale) as a function of the vibron number \(n\) for different coupling strengths \(\lambda>\Omega\). Increasing \(n\) leads to faster tunneling, with the growth in tunneling rate highest in the strong-coupling regime, showing near-exponential lifting of Franck-Condon blockade.}
\label{lindbladrates2}
\end{figure}

Next, we consider the situation where the vibron is prepared in a number state, resulting in the density matrix \(\rho_\text{vib}^{(n)}=\ket{n}\bra{n}\). We can obtain Lindblad coefficients $h^{(n)}_\text{in}$ and $h^{(n)}_\text{out}$ for any initial occupation number \(n\). The expressions are given in App.~\ref{sec_appcoeffn} since they are too lengthy to be shown here. In Fig.~\ref{lindbladrates} we show the decay rates as functions of the electron-vibron coupling \(\lambda/\Omega\). In contrast to the case of no initial vibrons, Eq.~\eqref{lowtemp}, we observe that a number of local maxima is superimposed onto the graphs, implying that the presence of vibrons eases the tunneling of an electron out of the dot. For coupling strengths beyond this non-monotonous region, a more pronounced increase of tunneling as a function of \(n\) is observed (see Fig.~\ref{lindbladrates2}), which becomes progressively closer to exponential as the coupling is made larger, implying that the conductive properties of the dot can be exponentially activated by exciting vibrons. Once more, this result is in full agreement with the rate-equation treatment from Ref.~[\onlinecite{koc06}], where the electron tunneling rates where found to be determined by the matrix elements
\begin{align}
M^{q\to q^{\prime}}_{1\to 0}=\left(\frac{\lambda}{\omega}\right)^{q-q^{\prime}}\mathrm{e}^{-\frac{\lambda^2}{2\omega^2}}\sqrt{\frac{q^{\prime}!}{q!}}L_{q^{\prime}}^{(q-q^{\prime})}\left(\frac{\lambda^2}{\omega^2}\right),
\end{align}
of transitions with initial and final vibron numbers \(q\) and \(q^\prime\leq q\), respectively.  Here \(L_n^{\alpha}(x)\) denote the generalized Laguerre polynomials. Since our approach does not resolve individual vibron transitions, the present results match \(\sum_{q^\prime=0}^q \vert M^{q\to q^\prime}_{1\to 0}\vert^2\), i.e., the sum of rates for all processes which either do not involve or de-excite vibrons.

\subsection{Vibron in a coherent state}
\label{sec_master3}

Finally, we move towards a vibron density matrix that is closer to the driven system we have in mind. A resonant drive of the electron level with amplitude \(A\) can be mapped to a vibron drive given by \(\mathrm{i}A'(a^\dagger\e^{-\ii\Omega_\text{dr}t}-a\e^{\ii\Omega_\text{dr}t})\), where \(A'=A\Omega/(2\lambda)\). If we in addition introduce a vibron damping rate \(\gamma\) to prevent divergences that might arise from driving a part of the bath, this generates the coherent vibron state\cite{baz13} \(\ket{\delta}\) with \(\delta\propto \mathrm{i}A'/(\Omega_\text{dr}-\Omega+\mathrm{i}\gamma)\). Therefore, we consider the vibron density matrix \(\rho_\text{vib}^{(\delta)}=\ket{\delta}\bra{\delta}\). For this case, we obtain the Lindblad coefficients
\begin{align*}
h^{(\delta)}_\text{in}=& \frac{\Gamma}{2}\e^{-\frac{\lambda^2}{\Omega^2}}\e^{-\mathrm{i}2\frac{\lambda}{\Omega}\left(\sin(\Omega t)\Re{\delta}-\cos(\Omega t)\Im{\delta}\right)}\\
&\times\sum_{m,l\in\mathbb{Z}}\e^{\mathrm{i}(m+l)\Omega t}\sum_{n\geq0}\frac{\left(\lambda^2/\Omega^2\right)^{n}}{n!}\mathrm{i}^lJ_m\left(2\frac{\lambda}{\Omega}\Re{\delta}\right)\\
&\times J_l\left(-2\frac{\lambda}{\Omega}\Im{\delta}\right)n_\text{F}(\tilde{\epsilon}+(m+l+n)\Omega), \\
h^{(\delta)}_\text{out}=& \frac{\Gamma}{2}\e^{-\frac{\lambda^2}{\Omega^2}}\e^{-\mathrm{i}2\frac{\lambda}{\Omega}\left(\sin(\Omega t)\Re{\delta}-\cos(\Omega t)\Im{\delta}\right)}\\
&\times\sum_{m,l\in\mathbb{Z}}\e^{\mathrm{i}(m+l)\Omega t}\sum_{n\geq0}\frac{\left(\lambda^2/\Omega^2\right)^{n}}{n!}\mathrm{i}^lJ_m\left(2\frac{\lambda}{\Omega}\Re{\delta}\right)\\
&\times J_l\left(-2\frac{\lambda}{\Omega}\Im{\delta}\right)\left(1-n_\text{F}(\tilde{\epsilon}+(m+l-n)\Omega)\right) \, . \numb
\end{align*}
Most strikingly, these tunneling rates are time dependent, reflecting the fact that coherent states are not eigenstates of the quadratic Hamiltonian \(\cH_0\). The influence of the drive on the decay rates can now be examined by varying the coherent state parameter \(\delta\). Specifically, the coherent state displacement \(\delta\) is proportional to the drive amplitude \(A'\), meaning it can be used as a measure of drive strength. Moreover, the squared absolute value \(|\delta|^2\) is proportional to the average number of vibrons in the coherent state, which establishes a link to the Fock state situation discussed previously.

In case of resonant driving, the real part \(\Re\delta\) dominates and its magnitude is directly proportional to the drive amplitude. However, the off-resonant situation can be studied in the same fashion by allowing for an imaginary part in \(\delta\). There it turns out that detuning between drive and vibron mode reduces the electron tunneling rate, which is in line with the intuition that the conductance is primarily affected by the number of vibrons present on the dot.

\begin{figure}[t]
     \includegraphics[width=0.99\columnwidth]{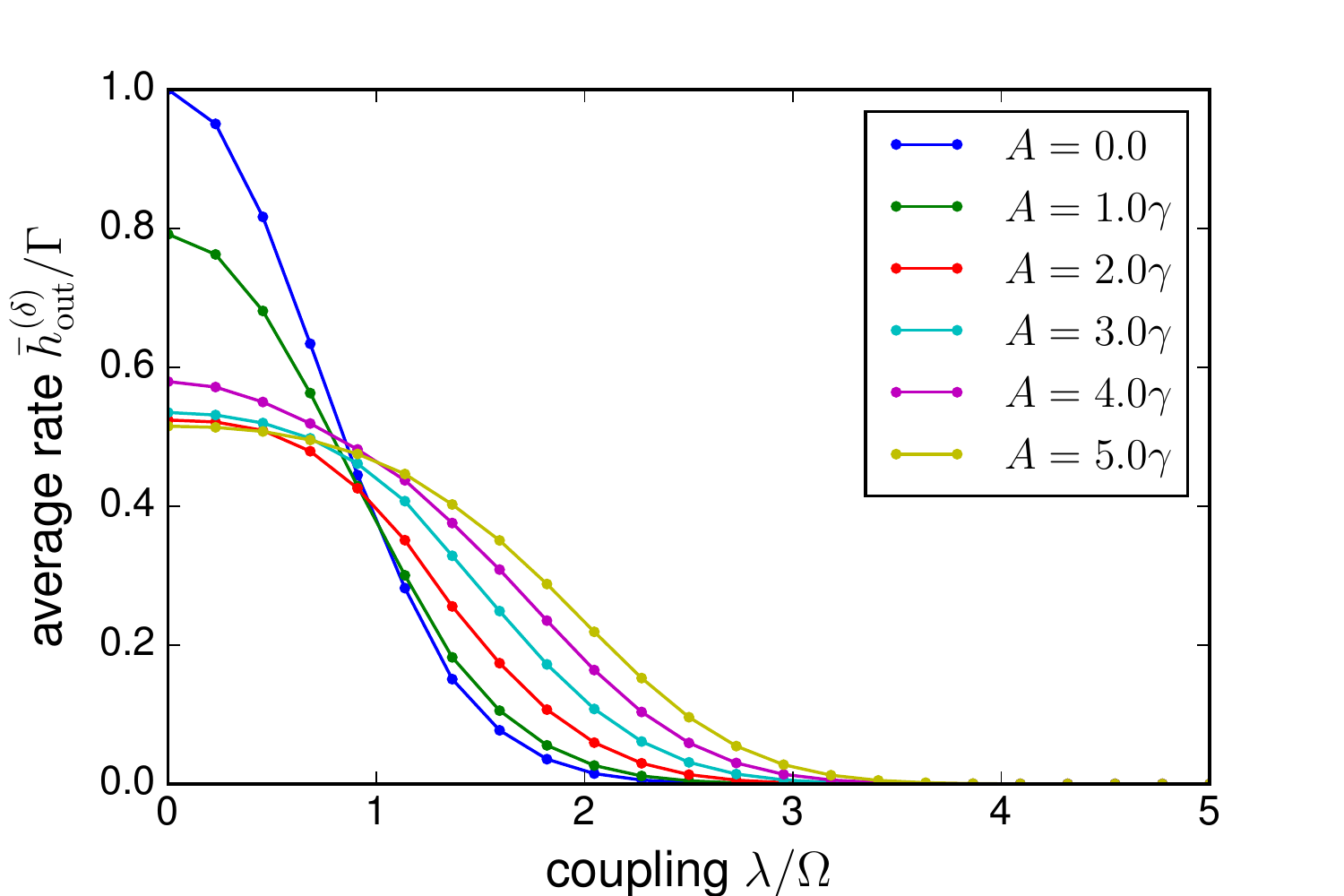}
  \caption{Dependence of the electron tunneling rate from the dot (at energy \(\tilde{\epsilon}=\Omega/2\)) into the lead (at zero energy) on the electron vibron coupling, obtained in the Lindblad formalism with a coherent vibron state. In general, the rate decreases strongly as a function of electron-vibron coupling, exemplifying Franck-Condon blockade. Two different regimes are apparent in the graphs: for weak electron-vibron coupling, the tunneling rate decreases with increasing drive amplitude \(A\) as a consequence of the oscillating dot energy dipping below the lead energy for increasing amounts of time. This decrease saturates for large drive, since then the level spends about half of a drive period below the lead energy. For stronger coupling, an increase in tunneling rates with \(A\) can be observed, the relative magnitude of which is moderate for \(1\lesssim\lambda/\Omega\lesssim3\), but becomes more substantial as the coupling is increased beyond that regime, see Fig.~\ref{cohlogrates}.}
\label{coh_plot}
\end{figure}

In order to make a comparison to the results obtained previously, we proceed to examine the average of \(h^{(\delta)}_\text{out}\) over one drive period, which is analogous to the averaging performed to obtain Eq.~\eqref{lambda2} in Sec.~\ref{sec_greens}. The outcome of this procedure is shown in Fig.~\ref{coh_plot}.

\begin{figure}[t]
     \includegraphics[width=0.99\columnwidth]{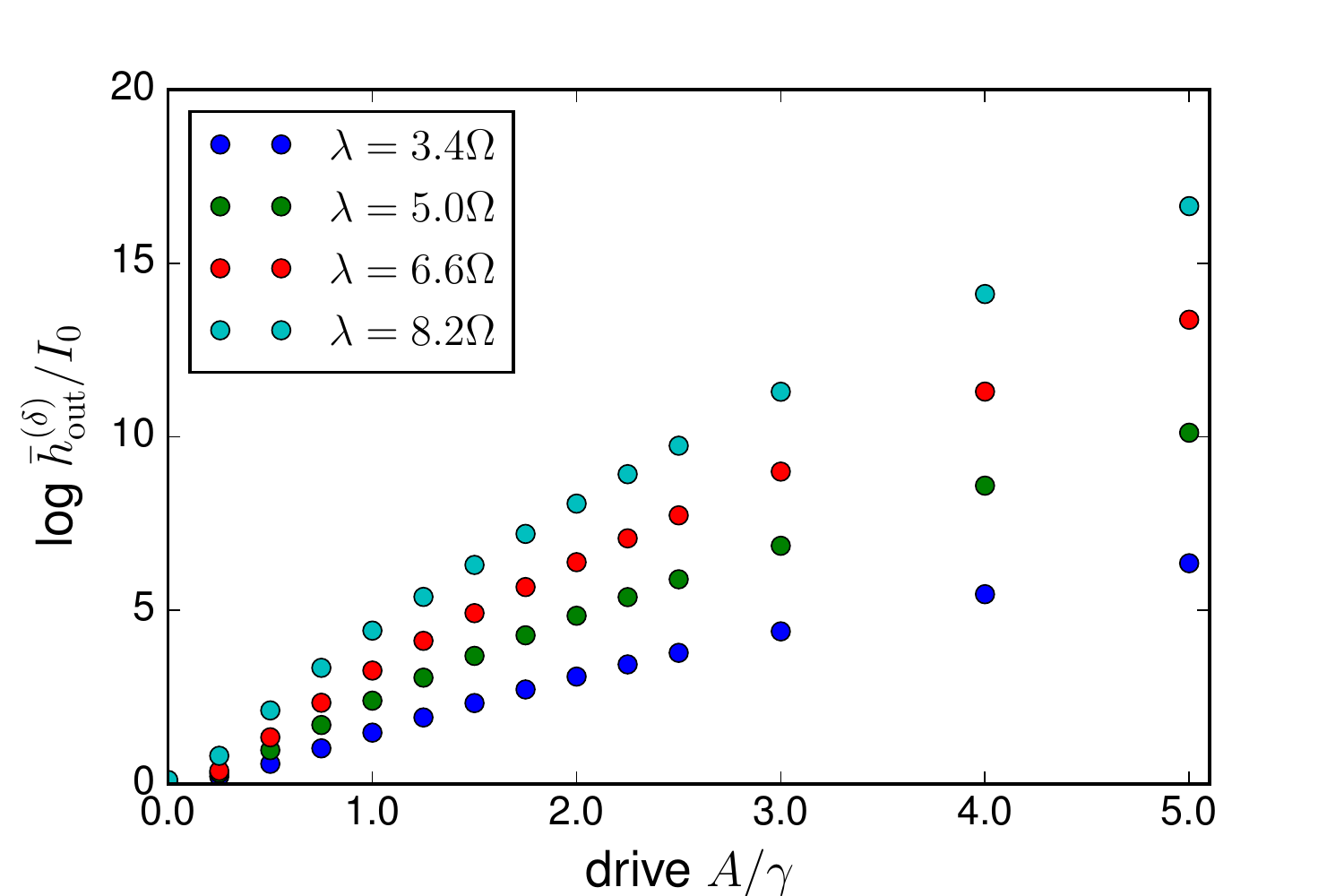}
  \caption{Tunneling rate (logarithmic scale) as a function of the drive amplitude \(A\), scaled by the inverse vibron damping \(\gamma\), for different coupling strengths \(\lambda>\Omega\). Increasing \(A\) leads to faster tunneling, with the growth in tunneling rate highest in the strong-coupling regime, exhibiting an exponential response over wide ranges of \(A\).}
\label{cohlogrates}
\end{figure}
Comparing the rates for different values of the drive amplitude uncovers three different regimes: $(i)$ a weak-coupling regime, where increasing the driving strength decreases the tunneling rates, $(ii)$ an intermediate regime showing a moderate increase of the tunneling rates for an increased  drive, and $(iii)$ a strong-coupling regime which features an almost exponential rise in tunneling rates as the driving strength is increased. The weak-coupling behavior is mainly an electronic effect: the favored tunneling transition here is the one that does not involve any vibrations. If the drive amplitude is less than the energy gap between dot and lead, there is little effect, but if it is significantly larger, the dot will spend a sizable part of the drive period below the lead; in this position, tunneling is energetically suppressed. For larger drive amplitudes, the portion of a drive period spent below the lead approaches \(1/2\), just like the resulting tunneling amplitude. For stronger coupling, the converse is true: driving the system makes it more conductive by activating the vibron-assisted tunneling channels. The intermediate regime is less extended here than in the case of vibrons in a Fock state, since \(\left\vert\delta\right\vert^2\), and hence the expected vibron number associated with the coherent state, decreases as \(\lambda^{-2}\). Lastly, the strong-coupling regime features the weakest currents, but also the strongest relative increase of \(h^{(\delta)}_\text{out}\) as a function of the driving strength, see Fig.~\ref{cohlogrates}, since conduction there involves the vibrations substantially, similarly to the Fock state case. This results in a current response that is almost perfectly exponential.
\begin{figure}[!t]
	\includegraphics[width=0.99\columnwidth]{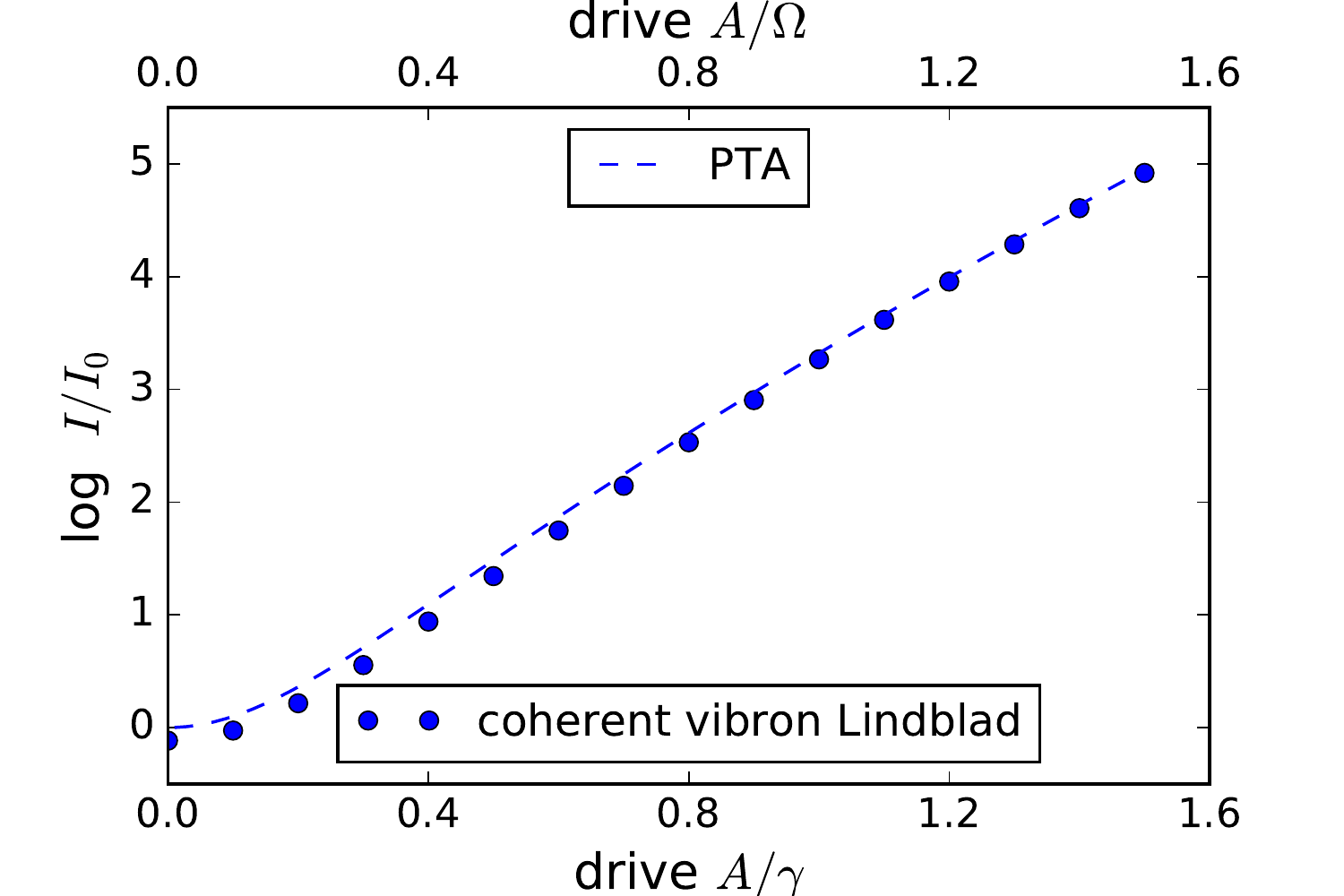}
\caption{Drive dependence (logarithmic scale) of the current through the resonantly driven quantum dot for the case of strong coupling, \(\lambda=6.6\Omega\), as derived in the Green's function and master equation formalisms, respectively. Both cases show a slightly subexponential response to driving, where the deviations from exponential behavior are more pronounced in the Green's function results. There, drive is parametrized by the amplitude \(A\), and in the master equation approach by the rescaling of \(A\) by the vibron damping rate.}
\label{logrates}
\end{figure}

The strong-coupling results admit a comparison to the findings of section \ref{sec_greens}, where we also derived a relation between the steady-state current and the applied drive strength. In Fig.~\ref{logrates} we compare the current obtained by both methods. We see satisfactory agreement between the two approaches in that, starting from the value of the Franck-Condon factor \(\e^{-\lambda^2/\Omega^2}\) in the undriven situation, both currents increase roughly exponentially as the drive is turned on. In both cases, further increase of the drive eventually shows a slight attenuation of current growth, resulting in deviations from an exponential characteristic. This leads us to conjecture that these deviations are of a physical nature and not just shortcomings of the specific method used.

\section{Conclusion}
\label{sec_discussion}
We studied the nonequilibrium behavior of a quantum dot with strong interaction between electronic and vibrational degrees of freedom, coupled to a pair of metallic leads in the sequential-tunneling regime. Using Keldysh Green's functions and a partial resummation of diagrams, we obtained the prediction that a periodically modulated gate voltage can be employed to change the transport properties of the system. Specifically, such a form of ac drive gives rise to an increase in steady-state current, lifting the Franck-Condon blockade. This finding turned out to be generally in line with the outcome of a master equation analysis of the electronic dynamics of system: both approaches show a strong current response to the gate voltage, which is close to exponential in the strongly interacting, weakly driven limit. The compatibility of these two results also supports the validity of the polaron-tunneling approximation used in the Green's function treatment in the presence of drive.

The results established here imply a transistor-like behavior of the CNT quantum dot, i.e., a conductance between a pair of leads that can be strongly activated by a gate voltage. The back gates that could be used to supply such a driving gate voltage are already part of some experimentally realized setups, in which they also have been used to modify the electron-vibron interaction and the coupling between quantum dot and leads. Hence, this points towards a variety of arrangements employing CNTs as electronic components of adjustable conductance, since both the coupling strengths and the drive parameters are in the range of experimental feasibility.

Challenges, however, are posed by the fact that the overall currents are still rather small, in spite of the exponential increase. It would therefore be worthwhile to study the intermediate regime where the electron-vibron coupling is not much larger than the frequency of the vibron mode. There, the conductivity is less strongly suppressed due to the Franck-Condon blockade, albeit at the price of more significant deviations from exponential response.

A further modification could be realized by incorporating the quantum dot into a setup related to circuit quantum electrodynamics (cQED),\cite{liu14,vie15} thus replacing the drive from the gate voltage with one generated by a microwave cavity.
Even more, cQED setups would also allow for replacing the classical drive (i.e.
a microwave cavity with a large number of photons) by a quantum drive.
In such configurations, measurements of the current flowing through the
CNT quantum dot could be used to as a detection mechanism for cavity photons.
Our finding that the conductance of the electromechanical component is strongly
actuated by drive suggests an arrangement of this kind as a possible high-precision
measurement device.

\begin{acknowledgments}
SW acknowledges financial support by the Marie Curie ITN cQOM and the ERC OPTOMECH. PH and TLS are supported by the National Research Fund Luxembourg \mbox{(ATTRACT 7556175)}. AN holds a University Research Fellowship from the Royal Society and acknowledges support from the Winton Programme for the Physics of Sustainability.
\end{acknowledgments}

\begin{widetext}
\appendix
\section{Details on the Keldysh Green's function approach}
\label{sec_app}

Below, we document the steps leading to  the renormalized dot Green's function \(D(\omega)\) which is used to obtain the current through the driven quantum dot in the Keldysh formalism.
The starting point is the polaron-transformed dot Green's function without renormalization due  to tunneling. Using the same notation for the vibrational operator \(\bar{X}\) as in the main text, this Green's function reads
\begin{align}
D_0(t,t^{\prime})=\braket{d^{\dagger}(t)\bar{X}(t)\bar{X}^{\dagger}(t^\prime)d(t^\prime)}_0. 
\end{align}
Since the expectation value is taken with respect to the ground state of the quadratic Hamiltonian \(\bar{\cH}_0=\Omega a^\dagger a+\tilde{\epsilon}d^\dagger d\), it factors into vibrational and electronic degrees of freedom, where the electronic component can immediately be written as 
\begin{align}
&G_{\text{dot}}(t-t^\prime)=-\mathrm{i}\mathrm{e}^{-\mathrm{i}\tilde{\epsilon}(t-t^{\prime})}\begin{pmatrix}-n_{d}+\Theta(t-t^{\prime})&&-n_{d}\\1-n_{d}&&-n_{d}+\Theta(t^{\prime}-t)\end{pmatrix},
\end{align}
with the Heaviside step function \(\Theta\), and initial dot occupation probability \(n_{d}\). For the vibron part, we find one of the Keldysh components to be
\begin{align*}
\Lambda^{+-}(t,t^\prime)=\braket{\bar{X}(t)\bar{X}^{\dagger}(t^{\prime})}_0
&=\left\langle\mathrm{e}^{-\frac{\lambda}{\Omega}\left[a^{\dagger}\mathrm{e}^{\mathrm{i}\Omega t}-a\mathrm{e}^{-\mathrm{i}\Omega t}+\mathrm{i} F(t)\right]}\mathrm{e}^{\frac{\lambda}{\Omega}\left[a^{\dagger}\mathrm{e}^{\mathrm{i}\Omega t^\prime}-a\mathrm{e}^{-\mathrm{i}\Omega t^\prime}+\mathrm{i} F(t^\prime)\right]}\right\rangle_0\\
&=\mathrm{e}^{-\frac{\lambda^2}{\Omega^2}}\mathrm{e}^{\frac{\lambda^2}{\Omega^2}\mathrm{e}^{-\mathrm{i}\Omega (t-t^{\prime})}}\mathrm{e}^{-\mathrm{i}\int_{t^{\prime}}^{t}\mathrm{d} sf(s)},\numb
\end{align*}
and analogous results for the others. Here the two exponentials on the left describe the undriven case, in particular the Franck-Condon blockade, and the drive is captured by the remaining factor, which depends on both the initial and final times \(t\) and \(t^\prime\). Since we want to capture the steady-state current through the system, we perform an average over one drive period. For the case of resonant driving, this rests upon the vibron timescale being much shorter than the measurement time. Thus we introduce the respective average and relative times \(T=(t+t^\prime)/2\) and \(\tau=t-t^\prime\), and proceed,
\begin{align*}
\label{lambda2calc}
&\frac{2\pi}{\Omega }\int_{-\frac{\pi}{\Omega }}^{\frac{\pi}{\Omega }}\mathrm{d} T\braket{\bar{X}(T+\tau/2)\bar{X}^{\dagger}(T-\tau/2)}_0
=\mathrm{e}^{-\frac{\lambda^2}{\Omega^2}}\mathrm{e}^{-\frac{\lambda^2}{\Omega^2}\mathrm{e}^{-\mathrm{i}\Omega \tau}}\frac{2\pi}{\Omega }\int_{-\frac{\pi}{\Omega }}^{\frac{\pi}{\Omega }}\mathrm{d} T\mathrm{e}^{-\mathrm{i}\int_{T-\tau/2}^{T+\tau/2}\mathrm{d} sf(s)}\\
=&\mathrm{e}^{-\frac{\lambda^2}{\Omega^2}}\mathrm{e}^{\frac{\lambda^2}{\Omega^2}\mathrm{e}^{-\mathrm{i}\Omega \tau}}\frac{2\pi}{\Omega }\int_{-\frac{\pi}{\Omega }}^{\frac{\pi}{\Omega }}\mathrm{d} T\mathrm{e}^{-\frac{iA}{\Omega }\left[\sin{\Omega \left(T+\tau/2\right)}-\sin{\Omega \left(T-\tau/2\right)}\right]}\\
=&\mathrm{e}^{-\frac{\lambda^2}{\Omega^2}}\mathrm{e}^{\frac{\lambda^2}{\Omega^2}\mathrm{e}^{-\mathrm{i}\Omega \tau}}\frac{2\pi}{\Omega }\int_{-\frac{\pi}{\Omega }}^{\frac{\pi}{\Omega }}\mathrm{d} T\sum_{n\in\mathbb{Z}}\mathrm{i}^{n}J_{n}\left(-\frac{2A\sin{\Omega \tau/2}}{\Omega }\right)\mathrm{e}^{\mathrm{i} n\Omega T}\\
=&\mathrm{e}^{-\frac{\lambda^2}{\Omega^2}\left(1-\mathrm{e}^{-\mathrm{i}\Omega\tau}\right)}J_0\left(\frac{2A\sin{\Omega \tau/2}}{\Omega }\right),\numb
\end{align*}
where \(f(t)=A\cos(\Omega t)\) denotes resonant drive.

The bare dot Green's function \(D_0\) is connected to the full PTA Green's function \(D\) via the frequency-space Dyson equation
\begin{align}
\label{adyson}
D^{-1}(\omega)=D_0^{-1}(\omega)-\Sigma_{\text{leads}}(\omega),
\end{align}
which has us perform the Fourier transform of \(D_0\). For the vibron degree of freedom, we note that the undriven part of the result in Eq.~\eqref{lambda2calc} can be immediately expanded into harmonics of \(\Omega\), whereas for the driven part, we can use the series expansion of the zeroth Bessel function, \(J_0(x)=\sum_{m\in\mathbb{N}}\frac{(-1)^m}{m!m!}\left(\frac{x}{2}\right)^{2m},\) to calculate the \(n\)th Fourier coefficient,
\begin{align*}
\label{driveft}
&\Lambda_{\text{dr}}^{(n)}=J_0\left(\frac{2A\sin\Omega t/2}{\Omega}\right)^{(n)}
=\frac{\Omega}{2\pi}\sum_{m\geq 0}\frac{(-1)^m}{m!m!}\frac{A^{2m}}{(\mathrm{i}2\Omega)^{2m}}\sum_{k=0}^{m}\begin{pmatrix}2m\\k\end{pmatrix}\int_{-\frac{\pi}{\Omega}}^{\frac{\pi}{\Omega}}\mathrm{d} t\mathrm{e}^{-\mathrm{i}(m-k)\Omega t}\mathrm{e}^{-\mathrm{i} n\Omega t}(-1)^{2m-k}\\
=&\sum_{m\geq |n|}\frac{(-1)^{m-n}}{m!m!}\left(\frac{A}{2\Omega}\right)^{2m}\begin{pmatrix}2m\\m+n\end{pmatrix} .\numb
\end{align*}

The dot Green's function is then transformed by convolving the above expression with the undriven vibron and bare electron parts,
\begin{align*}
D_0(\omega)=&\mathcal{F}\left[-\mathrm{i}\mathrm{e}^{-\mathrm{i}\tilde{\epsilon} t}\mathrm{e}^{-\frac{\lambda^2}{\Omega^2}}J_0\left(\frac{2A\sin(\Omega t/2)}{\Omega}\right)\right.\left.\begin{pmatrix}\Theta(t)\mathrm{e}^{\frac{\lambda^2}{\Omega^2}\mathrm{e}^{-\mathrm{i}\Omega|t|}}&0\\0&+\Theta(-t)\mathrm{e}^{\frac{\lambda^2}{\Omega^2}\mathrm{e}^{-\mathrm{i}\Omega|t|}}\end{pmatrix}\right](\omega)\\
=&-\mathrm{i}2\pi\mathrm{e}^{-\frac{\lambda^2}{\Omega^2}}\mathcal{F}\left[\sum_{n\in\mathbb{N}}\Lambda_{\text{dr}}^{(n)}\mathrm{e}^{\mathrm{i} n\Omega t}\sum_{k\in\mathbb{N}}\frac{1}{k!}\left(\frac{\lambda^2}{\Omega^2}\right)^k\mathrm{e}^{\mathrm{i} k\Omega t}\right.\left.\begin{pmatrix}\Theta(t)&0\\0&\Theta(-t)\end{pmatrix}\right](\omega-\tilde{\epsilon})\\
=&-\mathrm{i}2\pi\mathrm{e}^{-\frac{\lambda^2}{\Omega^2}}\sum_{n, k\in\mathbb{N}}\Lambda_{\text{dr}}^{(n)}\frac{1}{k!}\left(\frac{\lambda^2}{\Omega^2}\right)^k\begin{pmatrix}\int_{0}^{\infty}\mathrm{d} t\mathrm{e}^{\mathrm{i}\left( k\Omega +n\Omega+\omega-\tilde{\epsilon}\right)t}&0\\0&\int_{-\infty}^{0}\mathrm{d} t\mathrm{e}^{\mathrm{i}\left( k\Omega +n\Omega+\omega-\tilde{\epsilon}\right)t}\end{pmatrix}\\
=&2\pi\mathrm{e}^{-\frac{\lambda^2}{\Omega^2}}\lim_{\eta\to0}\sum_{n, k\in\mathbb{N}}\Lambda_{\text{dr}}^{(n)}\frac{1}{k!}\left(\frac{\lambda^2}{\Omega^2}\right)^k\begin{pmatrix}\frac{1}{k\Omega +n\Omega+\omega-\tilde{\epsilon}+\mathrm{i}\eta}&0\\0&-\frac{1}{k\Omega +n\Omega+\omega-\tilde{\epsilon}-\mathrm{i}\eta}\end{pmatrix},
\numb
\end{align*}
where \(\mathcal{F}\) denotes Fourier transform between time and frequency domains, and the cutoff \(\eta\to0^+\) serves to keep the integrals finite. We dropped the delta functions on the off-diagonal of the Fourier-transformed bare dot electron Green's function because those will not contribute when we invert the matrix in the next step.

Similarly to the bare dot electron Green's function, the lead Green's function takes shape as
\begin{align}
G_{\text{leads}}(\omega)=\sum_{k}\begin{pmatrix}\mathrm{i}2\pi n_{\text{F}}(\omega_k)\delta(-\omega_k+\omega)+\frac{1}{-\omega_k+\omega+\mathrm{i}\eta}&\mathrm{i}2\pi n_{\text{F}}(\omega_k)\delta(-\omega_k+\omega)\\\mathrm{i}2\pi (n_{\text{F}}(\omega_k)-1)\delta(-\omega_k+\omega)&\mathrm{i}2\pi n_{\text{F}}(\omega_k)\delta(-\omega_k+\omega)-\frac{1}{-\omega_k+\omega-\mathrm{i}\eta}\end{pmatrix},
\end{align}
for each of the two leads, where \(n_\text{F}\) again denotes the lead Fermi distribution. Taking the wide-flat-band limit of the lead distribution, we substitute \(\sum_k\to\int\frac{\mathrm{d} k}{2\pi}\to\int\mathrm{d} E\nu(E)\approx\nu\int\mathrm{d} E\), which renders the integration trivial and leaves us us with
\begin{align}
G_{\text{leads}}(\omega)=\mathrm{i}2\pi\nu\begin{pmatrix}n_\text{L}+n_\text{R}-\frac{1}{2}&n_\text{L}+n_\text{R}\\n_\text{L}+n_\text{R}-1&n_\text{L}+n_\text{R}-\frac{1}{2}\end{pmatrix}.
\end{align}
In the polaron tunneling approximation, this Green's function gives rise to the self-energy \(\Sigma_\text{leads}=g^2G_\text{leads}\).

 Now we are in position to use Eq.~\eqref{adyson} to calculate the PTA dot Green's function,
\footnotesize
\begin{align*}
&D(\omega)=\left(D_{0}(\omega)^{-1}-\Sigma_{\text{leads}}(\omega)\right)^{-1}\\
=&\frac{1}{\det D^{-1}(\omega)}\begin{pmatrix}-\left[2\pi\mathrm{e}^{-\frac{\lambda^2}{\Omega^2}}\sum_{n\in\mathbb{Z}, k\in\mathbb{N}}\Lambda_{\text{dr}}^{(n)}\frac{1}{k!}\left(\frac{\lambda^2}{\Omega^2}\right)^k\frac{1}{k\Omega +n\Omega+\omega-\tilde{\epsilon}}\right]^{-1}-\mathrm{i}\Gamma\left(n_\text{L}+n_\text{R}-\frac{1}{2}\right)&\mathrm{i}\Gamma (n_\text{L}+n_\text{R})\\\mathrm{i}\Gamma(n_\text{L}+n_\text{R}-1)&\left[2\pi\mathrm{e}^{-\frac{\lambda^2}{\Omega^2}}\sum_{n\in\mathbb{Z}, k\in\mathbb{N}}\Lambda_{\text{dr}}^{(n)}\frac{1}{k!}\left(\frac{\lambda^2}{\Omega^2}\right)^k\frac{1}{k\Omega +n\Omega+\omega-\tilde{\epsilon}}\right]^{-1}-\mathrm{i}\Gamma\left(n_\text{L}+n_\text{R}-\frac{1}{2}\right)\end{pmatrix}\numb
\end{align*}
\normalsize
with the determinant of the inverse Green's function matrix given by
\begin{align}
\det D^{-1}(\omega)=-\left[2\pi\mathrm{e}^{-\frac{\lambda^2}{\Omega^2}}\sum_{n\in\mathbb{Z}, k\in\mathbb{N}}\Lambda_{\text{dr}}^{(n)}\frac{1}{k!}\left(\frac{\lambda^2}{\Omega^2}\right)^k\frac{1}{k\Omega +n\Omega+\omega-\tilde{\epsilon}}\right]^{-2}-\frac{\Gamma^2}{4}.
\end{align}
The sum in this expression runs over all resonances of the vibron mode, where each resonance peak is in turn dressed by drive-induced contributions of the other ones. The resonances are Poisson-weighted in the case of no drive (\(A=0\)), and the weakly driven case, where \(\Lambda^{(n)}\) is strongly localized around \(n=0\), may be seen as a perturbed version of this.

The analysis in the main text is concerned with the drive dependence of \(D(\omega)\) around the lowest resonance, which is obtained from the above result by taking \(n=-k\).
\normalsize

\section{Details on the tunneling rates for Fock vibron state}
\label{sec_appcoeffn}

Here, we provide the Lindblad coefficient for the situation of a vibron prepared in a Fock state, with density matrix \(\rho_\text{vib}^{(n)}=\ket{n}\bra{n}\). In order to cast the master equation Eq.~\eqref{rhoeq2} into Lindblad form, we calculate the vibron trace
\begin{align}
\bra{n}X^{\dagger}(t)X(t-s)\ket{n}=\e^{-\frac{\lambda^2}{\Omega^2}}\sum_{m\geq0}\frac{(\lambda^2/\Omega^2)^{m}}{m!}\sum_{k=0}^{n}\begin{pmatrix}n\\k\end{pmatrix}\frac{(-2\lambda^2/\Omega^2)^{k}}{k!}\sum_{l=0}^{k}\begin{pmatrix}k\\l\end{pmatrix}\frac{(-1)^{k-l}}{2^{k-l}}\sum_{p=0}^{k-l}\begin{pmatrix}k-l\\p\end{pmatrix}\e^{\ii\Omega s(2p-k+l-m)},
\end{align}
which gives rise to the Lindblad coefficients
\begin{align*}
h^{(n)}_\text{in}=& \frac{\Gamma}{2} \e^{-\frac{\lambda^2}{\Omega^2}}\sum_{m\geq0}\frac{(\lambda^2/\Omega^2)^{m}}{m!}\sum_{k=0}^{n}\begin{pmatrix}n\\k\end{pmatrix}\frac{(-2\lambda^2/\Omega^2)^{k}}{k!}\sum_{l=0}^{k}\begin{pmatrix}k\\l\end{pmatrix}\frac{(-1)^{k-l}}{2^{k-l}}\sum_{p=0}^{k-l}\begin{pmatrix}k-l\\p\end{pmatrix}n_\text{F}(\tilde{\epsilon}-(2p-k+l-m)\Omega),\\
h^{(n)}_\text{out}=&\frac{\Gamma}{2} \e^{-\frac{\lambda^2}{\Omega^2}}\sum_{m\geq0}\frac{(\lambda^2/\Omega^2)^{m}}{m!}\sum_{k=0}^{n}\begin{pmatrix}n\\k\end{pmatrix}\frac{(-2\lambda^2/\Omega^2)^{k}}{k!}\sum_{l=0}^{k}\begin{pmatrix}k\\l\end{pmatrix}\frac{(-1)^{k-l}}{2^{k-l}}\sum_{p=0}^{k-l}\begin{pmatrix}k-l\\p\end{pmatrix}\left(1-n_\text{F}(\tilde{\epsilon}-(2p-k+l+m)\Omega)\right).\numb
\end{align*}
The rates resulting from these coefficients are visualized in Fig.~\ref{lindbladrates} of the main text.
\end{widetext}

\bibliography{bibliography}

\end{document}